# A NEW TRIFOCAL CORNEAL INLAY FOR PRESBYOPIA

*Walter D. Furlan[1]\*, Diego Montagud-Martínez[2], Vicente Ferrando[2], Salvador García-Delpech[3] and Juan A. Monsoriu[2]*

\*Corresponding Author: E-mail: walter.furlan@uv.es

[1]Departmento de Óptica y Optometría y Ciencias de la Visión, Universitat de València, 46100, Spain.
[2]Centro de Tecnologías Físicas, Universitat Politècnica de València, Valencia, 46022,
[3]Ophthalmology Department, Hospital Universitario La Fe, Valencia, 46026, Spain.

## Abstract

Corneal inlays (CIs) are the most recent surgical procedure for the treatment of presbyopia in patients who want complete independence from the use of glasses or contact lenses. Although refractive surgery in presbyopic patients is mostly performed in combination with cataract surgery, when the implantation of an intraocular lens is not necessary, the option of CIs has the advantage of being minimally invasive. Current designs of CIs are, either: small aperture devices, or refractive devices, however, both methods do not have good performance simultaneously at intermediate and near distances in eyes that are unable to accommodate. In the present study, we propose the first design of a trifocal CI, allowing good vision, at the same time, at far, intermediate and near vision in presbyopic eyes. We first demonstrate the good performance of the new inlay in comparison with a commercially available CI by using optical design software. We next confirm experimentally the image forming capabilities of our proposal employing an adaptive optics based optical simulator. This new design also has a number of parameters that can be varied to make personalized trifocal CI, opening up a new avenue for the treatment of presbyopia.

## Introduction

Presbyopia is the most common refractive defect in the population, affecting the quality of life of people over 45 years old, which nowadays exceeds two billion people worldwide. [1]. Its treatment, aimed to restore the ability to see clearly objects at near distances (depleted by the loss of accommodation) has multiple options, including multifocal spectacles, contact lenses and refractive surgery. Within this last option, the most recent approach is the implantation of CIs, entailing a minimally invasive and reversible surgery [2,3]. Currently, all CIs are implanted monocularly in the nondominant eye, producing a variant of the monovision technique, that consists in using the dominant eye for distance vision and the nondominant one for intermediate-near vision. CIs are small devices of a biocompatible material that are implanted into 'pockets' in the corneal stroma created by cavitation using femtosecond lasers. Thus, special care must be taken in the design of these devices and/or in the choice of inlay material to avoid the interruption of the normal cell activity in the stroma around it.

Based on different physical principles, several types of CIs have been proposed, each one having its own strengths and weaknesses [2,3]. At present, the most successful, and widely studied, commercial CIs are the refractive Flexivue Microlens® (Presbia Cooperatief, UA, Irvine, CA, USA) [4-7], and the small aperture KAMRA® inlay (Acufocus, Inc., Irvine, CA, USA) [8-10].

Refractive Inlays (RI) act locally at central part of the cornea either, by modifying its curvature or by altering the refractive index to improve near vision. However, RIs produce a loss of uncorrected distance visual acuity (UDVA) and contrast sensitivity [4,5,7]. Besides, an increase of higher order aberrations in the operated eye, especially spherical aberration has also been reported [6, 11]. To avoid degenerative material deposition and inflammation, refractive non-porous inlays should be manufactured with materials that ensure that flux of metabolic species is not modified by the device [12].

On the other hand, small aperture corneal inlays are simply opaque discs with a central hole that, acting as pinhole, produces an extended depth of focus at the cost of a loss of contrast sensitivity in the image. Thousands of micro-holes are randomly distributed on its surface to allow the passage of nutrients. The main shortcomings of small aperture inlays are associated with the intrinsic low light-throughput of the quasi-opaque ring. In fact, as the amount of light that reaches the retina of the fellow eye is significantly higher, the binocular distance visual performance, and the stereoacuity for near and intermediate distances are adversely affected [13-15]. Moreover, the diffraction produced by these pores aggravates the loss of contrast sensitivity previously mentioned.

Diffractive corneal inlays (DCIs) are the latest reported phakic surgery for the presbyopia correction [16]. This proposal is still under development, but it was presented as a promising alternative to solve some of the abovementioned drawbacks of refractive and small aperture corneal inlays. As the name indicates, DCIs work under the physical principle of diffraction, and are based on the so-called "photon sieve" concept, which was first proposed by Kipp et al. [17] for focusing X-rays. In a photon sieve, the alternate transparent and opaque rings of an amplitude Fresnel zone plate (a binary diffractive lens) are replaced by an arrangement of non-overlapping micro-holes distributed in the corresponding transparent Fresnel zones. Several unique and interesting properties of photon sieves were exploited in different areas [18-20]. Recently, our group proposed the first DCI as a combination of the photon sieve and the small aperture corneal inlay concepts. A DCI is in practice an opaque ring with thousands of micro-holes in its surface that in addition to allow the flow of nutrients, are strategically allocated to produce a focal point meant to see at near distances, thus converting the cornea into a bifocal optical system [21]. Moreover, it has been proposed that by optimizing the size and spatial distribution of the micro-holes, different designs would be able to vary the addition and the intensity ratio of different focal spots can be controlled through adjusting the proportion of the area of the DCI central hole and the surrounding structure [22]. The performance of our bifocal diffractive inlay has been verified by numerical simulation and optical bench experiments [16,21-23]. In spite of their improved light transmission efficiency with respect to the small aperture corneal inlays, previous models of DCIs have still a low light throughput due to the high proportion of opaque area.

On the other side, contrary to (premium) trifocal intraocular lenses, which are nowadays a very well established alternative to bifocal and monofocal lenses in cataract surgery, CIs have not yet passed yet the bifocal era. Indeed, patients implanted with CIs frequently still need spectacles for near or intermediate clear vision.

In this paper we propose the first trifocal corneal inlay for the treatment of presbyopia, which additionally is a pure phase diffractive device that, opposite to previous amplitude diffractive proposals, is fully transparent to improve light efficiency. We assessed the image quality and optical properties of this device, named Phase Diffractive Corneal Inlay (PDCI), in comparison with those of a commercially available refractive CI. To this end, Zemax OpticStudio design software (version 18.7, LLC, Kirkland, WA, USA) was employed to simulate the effects of both inlays in the Liou-Brennan model eye [24]. In the analysis, we used the modulation transfer function (MTF), the area under the MTF (AMTF), as merit functions; and a visual optical simulator has been employed to obtain the images provided by both inlays of objects at different vergences.

## Results

The PDCI here presented is a diffractive lens constructed by micro-holes drilled in a single sheet of a pure phase biocompatible material, which as shown in Fig. 1, is intended to be implanted in the cornea of a presbyopic eye. The optical quality of the PDCI was evaluated comparatively with a commercially available Refractive Corneal Inlay (RCI); fist, numerically by using Zemax software, and later, experimentally with an adaptive optics visual simulator with an artificial eye.

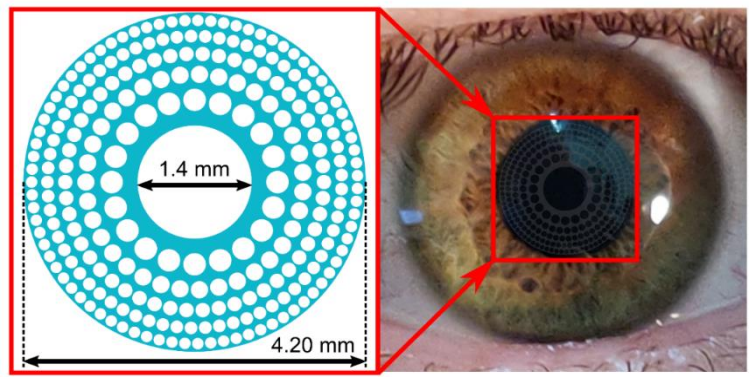

*Figure 1: PDCI design. The blue areas in the left figure represent a biocompatible (transparent), hydrogel-based, material of refractive index of 1.458 (see Methods section for details). The image on the right is a simulation of the appearance of the PDCI (in stark contrast) on a real eye.*

**Numerical results**

Figure 2a shows the trough-the-focus AMTFs, computed under polychromatic light, for both CIs in the Liu-Brennan model eye with 3.0 mm and 4.5 mm pupil diameters. The best far-distance focus was obtained for each device independently.

Continuous lines are the results obtained with the inlays centered on the visual axis. As can be seen, the PDCI presents a clear trifocal profile, with an intermediate focus at 1.75 D and a near focus at 3.0 D, which are maintained with both pupil diameters. On the contrary, as can be seen in the same figure the behavior of the RCI is very much pupil dependent. In fact, for a 3.0 mm pupil diameter, the RCI is clearly monofocal (near vision), but for 4.5 mm pupil it turns into a bifocal device with a higher value of the AMTF at the far focus.

To consider the influence of the CIs centration in the expected outcomes of the surgery, we have computed the AMTFs for the same pupil diameters but with the inlays decentered 1.0 mm towards the temporal direction in the model eye. Dotted lines in Fig. 2a show the results. As can be seen, the larger diameter of the PDCI, which is feasible because its high porosity that does not interfere with the passage of nutrients, is beneficial in making this device less sensitive to decentering than the RCI. Indeed, for a 3.0 mm diameter pupil, a decentering of 1.0 mm is already sufficient for some of the light to pass outside the RCI, which results in a very noticeable change between the centered and decentered AMTFs (see the blue lines in Figs. 2a).

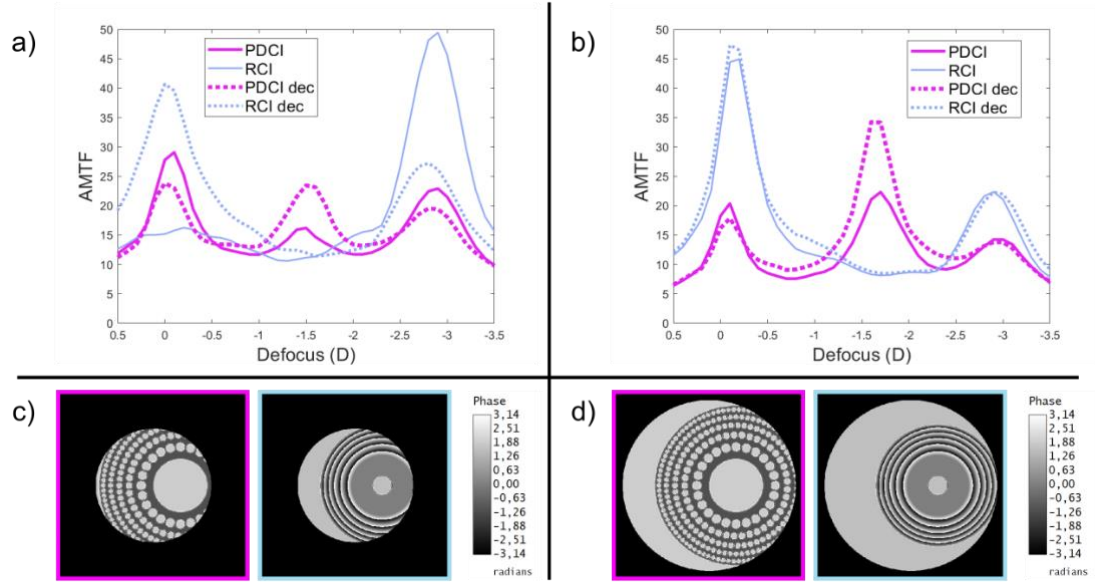

*Figure 2: AMTFs of PDCI (magenta) and RCI (blue) for 3.0 mm (a), and 4.5 mm pupil (b). Dotted lines correspond to the AMTFs with the optical inlays decentered 1.0 mm with respect to the pupil center as shown in the phase maps shown with the corresponding color frames in (c) for 3.0 mm and (d) for 4.5 mm pupil diameters.*

As, it has been recently shown that visual acuity (VA) defocus curves can be approximately predicted using a semiempirical non-linear function of the monochromatic (green light) AMTF [25], we have employed the mathematical expression reported in that work i.e:

$$\text{VA} = 5.06 \exp\left(\frac{\text{AMTFg}}{3.03}\right), \qquad (1)$$

to obtain the VA in logMAR units. In this expression the AMTFg is the monochromatic AMTF obtained for the wavelength reported in Ref. [25] (530 nm). The result for the 4.5 mm pupil diameter is shown in Fig. 3. Our aim was to compare this numerical result with the experimental results obtained with the adaptive optics visual simulator as will be shown in the next subsection.

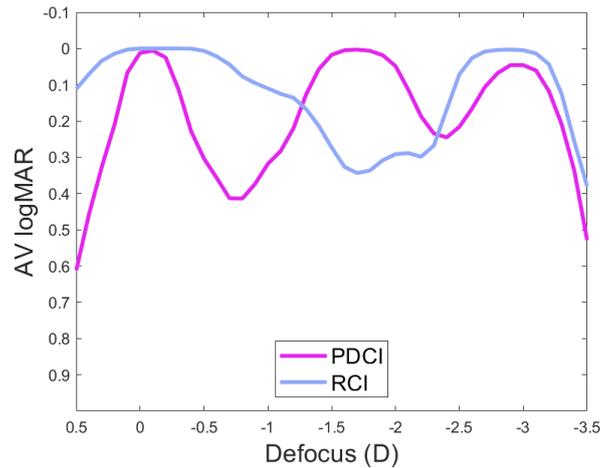

*Figure 3: Through-focus VA curves for the PDCI (magenta), and RCI (blue) obtained from the AMTFs using Eq. (1). The pupil size is 4.5 mm and the abscissa axis has the origin (0.0 D defocus) at the distance focus of each lens.*

**Experimental results**

An adaptive optics-based visual simulator (VAO system, VOptica, Murcia, Spain) was employed to get experimental images provided by the PDCI in comparison with those obtained with the RCI. This system allows simulating vision with any phase device virtually implanted in the tested eye with a pupil diameter of 4.5 mm and has been measured with different ophthalmic elements. The test object for the experiment was a tumbling E optotype with three different letter sizes corresponding to logMAR visual acuities of 0.4, 0.2 and 0.0. The recorded images of the optotype by a CCD camera (acting as an artificial eye) at different vergences are shown in Fig. 4. These images were taken using the green channel of the VAO system in order to correlate the results with the numerical simulations shown in Fig. 3. As can be seen, the VA images are in agreement with the theoretical predictions. In particular, for the PDCI images note the asymmetry in the depth of focus of the near and far foci. In fact, the curve in the Fig. 3 predicts a better image for -2.5 D than for -0.5D (despite both are 0.5 D apart from the near and far focus respectively). Interestingly, just the opposite happens for the RCI, i.e.; the image at -0.5D is better than the image at -2.5 D, which is in accordance with the blue curve in Fig. 3. Moreover, the absence of intermediate focus for the RCI and the lower contrast of the images, at far and near distances, obtained with the PDCI shown in Fig. 4 were also predicted in Fig 3.

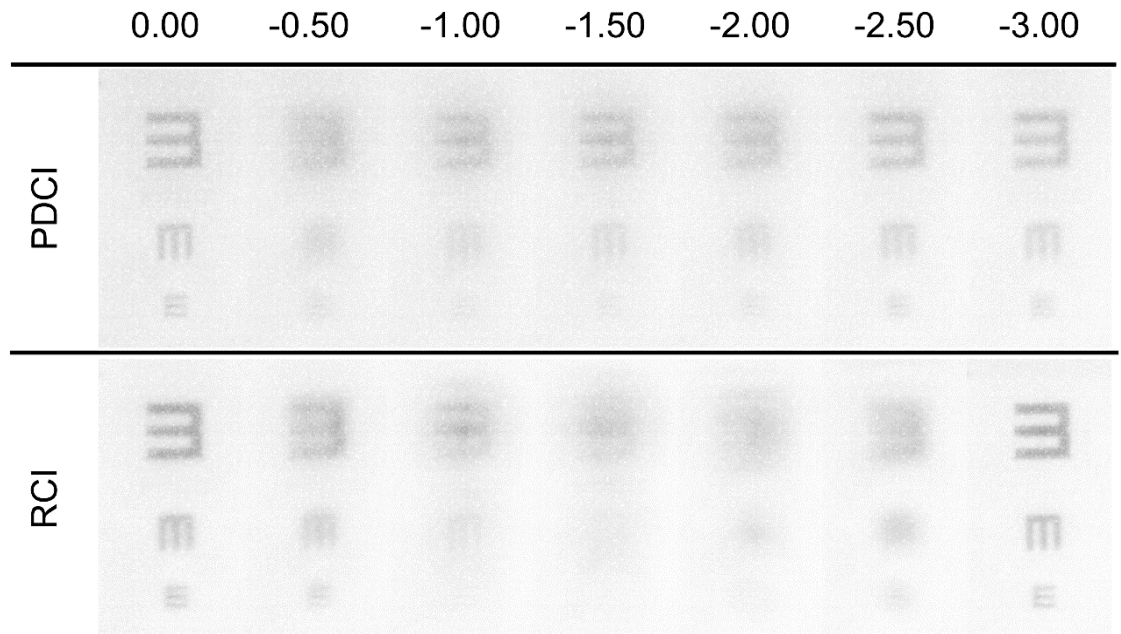

*Figure 4. Images of a tumbling E optotype corresponding to 0.4 ,0,2 and 0 logMAR VA obtained the VAO system simulating the PDCI and RCI with the object at different vergences from 0.0 D to -3.0 D.*

## Discussion and Conclusions

In this work, we have presented the design and optical properties of the first fully-transparent trifocal corneal inlay. This new device (PDCI) represents a considerable potential advantages over previous models of diffractive inlays [21-23], which on the one hand, are simply bifocals, and, on the other hand, have a lower luminous efficiency because they are partially opaque. We have demonstrated theoretically and experimentally that the PDCI presents good visual performance at intermediate distances while, at far and near distances the results are comparable to those obtained with a commercial RCI (Flexivue) under the same pupillary conditions. Prior to this, several studies have shown that such an inlay is clinically effective for the treatment of presbyopia [5,7,11]. However, it was found that the gain in near visual acuity in the operated eye is always accompanied by a loss of distance visual acuity, thus monovision is mandatory with this inlay model to preserve good binocular distance visual acuity with some independence on pupil size. In this sense, another result of this work was the assessment of the influence of different pupil diameters on the vision of objects at different distances for the virtually operated eye with both inlays [Fig.2]. It is important to mention here that clinical studies of Flexivue, reported VA outcomes but do not mention the pupillary conditions under which they were measured. Another essential point to be considered in the outcomes of CI surgery is the centering of the implant [11, 26]. However, for RCIs there are no quantitative results to justify this hypothesis. Currently only two studies [4,11] describe only qualitatively what could happen with a Flexivue decentering. In this work the offset of the inlays was numerically evaluated and it is proven that it is critical for RCI in small pupils (See Fig. 2a.). Thus, the results here presented provide additional information about the RCIs not reported previously, and highlight the importance of both, pupil size and centering. Importantly, we have predicted that our proposal is more robust than the RCI in both aspects.

As a common feature with other corneal surgeries, our proposal could be practiced concurrently or independently with LASIK or PRK in myopic or hyperopic eyes [27]. However, just as the optimum candidates for small aperture corneal inlays, are slightly myopic eyes [28], in our case this optimum condition would be obtained for slightly hyperopic eyes to take advantage of the virtual focus for its use as a far distance focus. Related to this previous refractive condition, as the kappa angle depends on the axial length, with hyperopic eyes tending to have a larger angle kappa than myopic eyes [28], the robustness against decentering can be considered another important property of the PDCI. Complimentary to this, and advantage in our case is that there is a certain degree of freedom to design the implant so that the value of the near (and intermediate) addition could be varied. Moreover, for a given value of the addition, the spatial distribution and diameters of micro-holes in each zone can also be modified to obtain an optimized relative intensity between the near and far foci. Even more, further improvements in customized PDCIs models could are feasible considering that the micro-holes density along the radial and azimuthal coordinates can be varied to achieve sphero-cylindrical PDCI for astigmatism and/or high order ocular aberrations.

Finally, it is important to note that the photon sieve concept applied to CI designs opens other interesting options to be explored in the future —some of which are already under way. These include the use of other multifocal diffractive structures (fractal [29], Fibonacci [30], Thue-Morse [31], etc.); and, remarkably, taking into account that the flux of metabolic species is affected by any inlay (especially those made of non-porous materials) the effect of the inlay on the long-term health of the cornea is of primary importance. In this sense, our design is fully compatible with the recent advances reported in 3D bioprinting of corneal stroma equivalents with highly transparent, biocompatible, and stable materials. [32].

In conclusion, in this study, we have demonstrated the feasibility of the first trifocal corneal inlay for the presbyopia treatment. Our proposal was numerically and experimentally, evaluated in comparison with a commercially available refractive corneal inlay. Trifocality, and robustness against decentration are benefits not previously reported simultaneously by any other CI. Thus, the implantation of PDCI seems to be an interesting alternative to be explored for phakic presbyopic patients who desire spectacle independence, and would be fully compatible with (previous or in combination) laser refractive procedure in myopic and hyperopic patients; and also, with cataract surgery afterwards.

## Methods
### Lens design and characteristics

The PDCI is the evolution of previous deigns of DCIs proposed by our group [21-23] in which we have combined two physical principles: the extended depth of focus provided by a mask with a small aperture (pin-hole), and the photon sieve. Therefore, previous DCIs designs are essentially opaque disks with a central hole and thousands of micro-holes distributed into annular zones that coincide with those of a Fresnel zone plate. By properly positioning and sizing of these micro-holes, only the zeroth and first positive and negative diffraction orders foci are present, and the high orders of the underlying conventional Fresnel zone plate are almost suppressed. It is very well known that the low diffraction efficiency of amplitude Fresnel zone plates can be improved up to a factor of 4, by replacing the opaque areas of the plate by a transparent phase-type material having the appropriate thickness to introduce a □ phase shift between alternate zones [33]. This is the main idea behind the new design of DCI here presented. Complementary to this, several parameters can regulate the focusing performance of a PDCI. A typical example is shown in Fig. 1. First, as our aim is to obtain a trifocal device, we need to partially restore the $0^{th}$ diffraction order to use it as the focus for the intermediate distances. This can be achieved simply by modifying the diameter of the central hole $H$. The micro-holes in the periphery (odd rings of the zone plate) produce two main additional foci, the negative and positive and $1^{st}$ diffraction orders, which are intended to far and near distance vision, respectively. In general terms, the number $N$ of micro-holes on each ring and the diameter ($d$), of the holes in the $i^{th}$ ring determine the total PDCI patterned area and therefore the PDCI diffraction efficiency. The diameters of the holes in a conventional photon sieve are usually expressed as a function of the ring width $w$, as $d=Kw$, where $K$ is a constant. So, there is a compromise between $N$ and $d$ in order to avoid overlapping between holes, preserving in this way the PDCI in a single structure.

The refractive index of the material chosen for the construction of the PDCI, $n_I$ regulates the exact thickness of the inlay $h$ for the design wavelength $\lambda_0$:

$$h = \frac{\lambda_0}{2 \cdot (n_I - n_c)} \qquad (2)$$

The PDCI under test shown in Fig. 1 was designed to provide a near diffractive focus corresponding to a nominal addition (near focus) of +3 D, to compare its performance with the commercial RCI Flexivue Microlens (Presbia, Irvine, CA, USA). Hence, in our simulations the material selected for the PDCI was an hydrogel with a refractive index of index $n_I = 1.458$. We assumed the refractive index of the corneal stroma is $n_C = 1.376$ corresponding to one employed in the Liou-Brennan eye model ($\lambda_0 = 555$ nm), Using Eq. (1) the thickness of the inlay results: $h$=3.5 microns.

In our case the diffractive structure was a disk of 4.2 mm diameter with a central hole of 0.7 mm diameter, surrounded by 5 rings conformed by a total of 253 holes of different size d obtained with *K*=1.62, being the smallest ones of 75 □m. The optical characterization of the PDCI was initially performed using Zemax OpticStudio design software (v. 18.7, LLC, Kirkland, WA, USA) in comparison with the abovementioned Flexivue Microlens. This RCI is a transparent, hydrogel-based, concave–convex disc made out of hydroxyethyl methacrylate and methyl methacrylate with a 3.2 mm diameter and ~15–20 µm thickness [1] The central 1.6 mm diameter is plano, and the annular peripheral zone in our simulations had an add power of +3.0 D. At the center of the disc, a 0.51 mm hole facilitates the transfer of nutrients into the cornea through the lens. [7]

Table 1 shows the parameters of Liou-Brennan's theoretical model eye, which reflects average biometrical data from a large group of individuals; incorporating a grin based model crystalline lens and the corresponding inlay in each case. The insertion of the CIs, in the model eye, was introduced as a "Grid Sag Surface", both at the same depth, 0.3 mm, from the anterior surface of the cornea.

The AMTFs of the two CIs were calculated for frequencies between 0 and 50 cycles/degree. These spatial frequencies that correspond to decimal visual acuities up to 1.6, were employed to calculate the average of sagittal and tangential MTFs at different vergences: from +0.50 D to -3.50 D in 0.10 D steps, and with two different pupil diameters: 3.0 and 4.5 mm, emulating photopic and mesopic conditions. In the simulations, we employed two different settings for the illumination: monochromatic light, matching the design wavelength (555 nm) and polychromatic light using the *photopic bright* setting of Zemax which uses five weighted wavelengths. The optimum target for the far distance focus was obtained independently for each device taking the best AMTF value as a quality criterion. In the calculations the ideal eye's pre-surgical refractive state for the PDCI resulted +1.75D while for the RI was emmetropia. This is equivalent to assuming that the inlay surgery was performed simultaneously with LASIK or PRK in patients who are not already at an optimal preoperative refraction, which is a common and safe clinical procedure with commercial corneal inlays [34].

**Table 1. Liou-Brennan model eye Zemax data sheet**
(*r* and *z* are radial and axial coordinates in the crystalline lens)

| Surface | Radius (mm) | Asphericity | Thickness (mm) | Refractive index |
|---|---|---|---|---|
| Anterior Cornea | 7.77 | -0.18 | 0.300 | 1.376 |
| Anterior corneal inlay | 7.77 | -0.18 | 0.0035 | 1.458 |
| Posterior corneal inlay | 7.77 | -0.18 | 0.2965 | 1.376 |
| Posterior Cornea | 6.40 | -0.60 | 3.160 | 1.336 |
| Iris | - | - | 0.000 | - |
| Anterior Lens | 12.40 | -0.94 | 1.590 | $1.368 + 0.049057\,z - 0.015427\,z^2 - 0.001978\,r^2$ |
| Lens | Infinity | - | 2.430 | $1.407 - 0.006605\,z^2 - 0.001978\,r^2$ |
| Posterior Lens | -8.10 | 0.96 | 16.260 | 1.336 |

**Adaptive Optics Visual Simulator**

The experimental measurements in this work were taken using the VAO adaptive optics visual simulator (Voptica S.L., Murcia, Spain). This clinical instrument allows to place optical stimuli different vergences through a micro display and to show to the patient its image through any optical phase profile [35, 36]. The stimulus was an optotype with three high-contrast letters (tumbling Es) of different sizes corresponding to visual acuities of 0.4; 0.2 and 0.0 logMAR units. In our case, we have incorporated the phase of both inlays into the system following the indications of the manufacturer as CSV files with 846x846 values covering a pupil of 4.5 mm diameter. In this study, through-focus images provided by the corneal inlays in the VAO system were recorded in the range 0.0D - 3.0 D in 0.25 D steps using a 8 bits CCD camera (Edmund-Optics with model EO-10012C Lite Edition) with a resolution of 3840 x 2748 pixels and CCD dimensions of 6.41 x 4.59 (mm). The focusing lens was an achromatic doublet with 30 mm focal (AC254-030-A-ML, Thorlabs Inc. Newton, NJ, USA). Therefore, by recording images the visual stimuli through the VAO system, with a CCD camera replacing the eye, our aim was to found an agreement with the numerical simulations of the through the focus performance of both inlays. This was done despite that the real size of the projected phase CIs could not be measured experimentally because according to the manufacturer a real image of them is projected into the pupil plane of the observer's eye, but the exact location of this plane is not specified. Consequently, although the instrument works with a single pupil diameter of 4,5 mm its projection over the artificial eye could have a magnification slightly different than 1.

*Data availability: Data that support the findings of this study are available from the corresponding authors upon reasonable request.*

**Acknowledgments**: This work was supported by the Generalitat Valenciana, Spain, under Grant PROMETEO/2019/048. D. Montagud–Martinez and V. Ferrando acknowledge the financial support from the Universitat Politècnica de València, Spain (fellowships FPI–2016 and PAID–10–18, respectively).

**References**

[1] Fricke, T. R. et al. Global prevalence of presbyopia and vision impairment from uncorrected presbyopia: systematic review, meta-analysis, and modelling. *Ophthalmol.* **125,** 1492–1499 (2018).

[2] Charman, W. N. Developments in the correction of presbyopia II: surgical approaches. *Ophthal. Physiol. Opt.* **34**, 397-426 (2014).

[3] Moarefi, M. A., Bafna, S., Wiley, W. A review of presbyopia treatment with corneal inlays. *Opthamol. Ther.* **6,** 55-65 (2017).

[4] Beer, S. M. C., et al. A 3-year follow-up study of a new corneal inlay: clinical results and outcomes. British J. Ophthalmol. **104**, 723-728 (2020).

[5] Limnopoulou, A. N., et al. Visual outcomes and safety of a refractive corneal inlay for presbyopia using femtosecond laser. *J. Refract. Surg.* **29,** 12-18 (2013).

[6] Garza, E.B., Gomez, S., Chayet, A. & Dishler, J. One-year safety and efficacy results of a hydrogel inlay to improve near vision in patients with emmetropic presbyopia. *J. Refract. Surg.* **29,** 166-172 (2013).

[7] Malandrini, A. et al. Bifocal refractive corneal inlay implantation to improve near vision in emmetropic presbyopic patients. *J. Cataract Refract. Surg.* **41**, 1962-1972 (2015).

[8] Waring G. O. Correction of presbyopia with a small aperture corneal inlay. *J. Refract. Surg.* **27,** 842-845 (2011).

[9] Vilupuru, S., Lin, L., Pepose, J. S. Comparison of contrast sensitivity and through focus in small-aperture inlay, accommodating intraocular lens, or multifocal intraocular lens subjects. *Am. J. Ophthalmol.* **160,** 150-162 (2015).

[10] Vukich, J.A. et al. Evaluation of the small-aperture intracorneal inlay: Three-year results from the cohort of the US Food and Drug Administration clinical trial. *J.* Cataract. *Refract. Surg.* **44,** 541-556 (2018).

[11] Beer, S. M. et al. One-year clinical outcomes of a corneal inlay for presbyopia. *Cornea* **36**, 816-820 (2017).

[12] Pinsky, P. M. Three-dimensional modeling of metabolic species transport in the cornea with a hydrogel intrastromal inlay. *Invest. Ophthalmol. Vis. Sci.* **55**, 3093-3106 (2014).

[13] Gilchrist, J. & Pardhan, S. Binocular contrast detection with unequal monocular illuminance. *Ophthal. Physl. Opt.* **7,** 373-377 (1987).

[14] Plainis, S. et al. Small aperture Monovision and the Pulfrich experience: Absence of neural adaptation effects. *PLoS One* **8,** e75987; 10.1371/journal. Pone.0075987 (2013).

[15] Castro, J. J., Ortiz, C., Jiménez, J. R., Ortiz-Peregrina, S. & Casares-López, M. Stereopsis simulating small-aperture corneal inlay and Monovision conditions. *J. Refract. Surg.* **34,** 482-488 (2018).

[16] Furlan, W. D. et al. Diffractive corneal inlay for presbyopia. *J. Biophotonics* **10**, 1110-1114 (2017).

[17] Kipp, L. et al. Sharper images by focusing soft X-rays with photon sieves. Nature **414**, 184-188 (2001).

[18] Andersen, G. Large optical photon sieve. *Opt. Lett.* **30**, 2976-2978 (2005).

[19] Menon, R., Gil, D., Barbastathis, G., & Smith, H. I. Photon-sieve lithography. *J. Opt. Soc. Am. A* **22**, 342-345 (2005).

[20] Giménez, F., Monsoriu, J. A., Furlan, W. D. & Pons, A. Fractal photon sieve. *Opt. Express* **14,** 11958-11963 (2006).

[21] Montagud-Martínez, D., Ferrando, V., Machado, F., Monsoriu, J. A. & Furlan, W. D. Imaging performance of a diffractive corneal inlay for presbyopia in a model eye. *IEEE Access* **7**, 163933 (2019).

[22] Montagud-Martínez, D., Ferrando,V., Monsoriu, J. A. & Furlan, W. D. Optical evaluation of new designs of multifocal diffractive corneal inlays. *J. Ophthalmol.* **2019**, 9382467 (2019).

[23] Montagud-Martínez, D., Ferrando, V., Monsoriu, J. A. & Furlan, W. D. Proposal of a new diffractive corneal inlay to improve near vision in a presbyopic eye. *Appl. Opt.* **59**, d54-d58 (2020).

[24] Liou, H. L. & Brennan, N. A. Anatomically accurate, finite model eye for optical modeling. *J. Opt. Soc. Am. A* **14**, 1684-1695 (1997).

[25] Vega, F. et al. Visual acuity of pseudophakic patients predicted from in-vitro measurements of intraocular lenses with different design. *Biomed. Opt. Express* **9**, 4893-4906 (2018).

[26] Han, G. et al. Refractive corneal inlay for presbyopia in emmetropic patients in Asia: 6-month clinical outcomes. *BMC Ophthalmol.* **19**, 66 (2019).

[27] Roger, F. et al. Corneal remodeling after implantation of a shape-changing inlay concurrent with myopic or hyperopic laser in situ keratomileusis. *J. Cataract. Refract. Surg.* **43**, 1443-1449 (2017).

[**28**] Tabernero, J. & Artal, P. Optical modeling of a corneal inlay in real eyes to increase depth of focus: optimum centration and residual defocus. *J.* Cataract *Refract. Surg.* **38**, 270-277 (2012).

[29] Monsoriu, J. A., Saavedra, G. & Furlan, W. D. Fractal zone plates with variable lacunarity. *Opt. Express* **12**, 4227-4234 (2004).

[30] Monsoriu, J. A. et al. Bifocal Fibonacci diffractive lenses. *IEEE Photon. J.* **5**, 3400106-3400106 (2013). .

[31] Ferrando, V., Giménez, F., Furlan, W. D. & Monsoriu, J. A. Bifractal focusing and imaging properties of Thue–Morse Zone Plates. *Opt. Express* **23**, 19846-19853 (2015).

[32] Bektas, C. K. & Hasirci, V. Cell loaded 3D bioprinted GelMA hydrogels for corneal stroma engineering. *Biomater. Sci.* **8**, 438-449 (2020).

[33] Kirz, J. Phase zone plates for x rays and the extreme uv. *J. Opt. Soc. Am.* **64**, 301-309 (1974).

[34] Moshirfar, M. et al. Retrospective comparison of visual outcomes after KAMRA corneal inlay implantation with simultaneous PRK or LASIK. *J. Refract. Surg.* **34**, 310-315 (2018).

[35] Manzanera, S., Prieto, P. M., Ayala, D. B., Lindacher, J. M. & Artal, P. Liquid crystal Adaptive Optics Visual Simulator: Application to testing and design of ophthalmic optical elements. *Opt. Express* **15**, 16177-16188 (2007).

[36] Hervella, L., Villegas, E. A., Robles, C. & Artal, P. Spherical Aberration Customization to Extend the Depth of Focus With a Clinical Adaptive Optics Visual Simulator. *J. Refract. Surg.* **36**, 223-229 (2020).

## Author Contributions

W.F. conceived and designed the study, analysed and interpreted the data, and wrote the manuscript. D.M-M. participated in the design of the study, performed the optical simulations, analysed and interpreted the data, V. F participated in the design of the study, programmed the CIs, and performed preliminary Matlab simulations. S. G-D interpreted the data, and helped to write the manuscript, J. Monsoriu contributed essential materials and analysis tools, and revised draft the manuscript. All authors were involved in the discussion of results, critical reading and final approval of the manuscript.

**Competing interests**
The authors declare no competing interests.